\documentclass[12pt]{article}
\usepackage{color}
\usepackage{amsmath}%\pdfoutput=1
\usepackage{cite}
\usepackage[pdftex]{graphicx}
\begin{document}    
\begin{flushright}
%September, 2023
\end{flushright}
\vspace*{1cm}

\renewcommand\thefootnote{\fnsymbol{footnote}}
\begin{center} 
  {\Large\bf $CP$ issues in the SM from a view point of spontaneous $CP$ violation}
\vspace*{1cm}

{\Large Daijiro Suematsu}\footnote[1] {professor emeritus, ~e-mail:
suematsu@hep.s.kanazawa-u.ac.jp}
\vspace*{0.5cm}\\

{\it Institute for Theoretical Physics, Kanazawa University, 
Kanazawa 920-1192, Japan}
\end{center}
\vspace*{1.5cm} 

\noindent
{\Large\bf Abstract}\\
The standard model (SM) has several issues related to the $CP$ violation which could
give clues to search physics beyond the SM. They are a $CP$ phase in the CKM matrix, 
the strong $CP$ problem, $CP$ phases in the PMNS matrix and $CP$ asymmetry 
in lepton number violating processes related to baryon number asymmetry. 
We consider a model which could give a unified explanation for them in a framework 
of spontaneous $CP$ violation. It is an extension of the SM with vector-like 
fermions and singlet scalars. In this model, they are explained by a common complex 
phase caused in the spontaneous $CP$ violation. We present concrete examples for 
them and also discuss some relevant phenomenology. 

\newpage
%%%%%%%%%%%%%%%%%%%%%%%%%%%%%%%%%%%
\setcounter{footnote}{0}
\renewcommand\thefootnote{\alph{footnote}}

\section{Introduction}
Origin of $CP$ violation in the quark sector of the standard 
model (SM) is considered to be given by complex Yukawa couplings \cite{km}. 
They fix up-type and down-type $3\times 3$ quark mass matrices ${\cal M}_u$ and 
${\cal M}_d$. A $CP$ phase appears in the CKM matrix by considering their mass eigenstates.
Since it is irrelevant to a $\theta$ parameter in the QCD sector \cite{theta},
the strong $CP$ problem \cite{strongcp} is caused. 
Although an experimental bound for the neutron electric dipole moment \cite{nedm}
requires $\bar\theta=\theta+{\rm arg}[\det({\cal M}_u{\cal M}_d)]<10^{-10}$ 
\cite{expnedm}, we cannot explain why irrelevant ones can realize such a small value. 
This problem is known to be solved by the axion \cite{axion,ksvz,dfsz} caused by
spontaneous breaking of the Peccei-Quinn (PQ) symmetry \cite{pq}.
Axion physics severely constrains a breaking scale of the PQ symmetry \cite{fa}. 

An alternative solution for the strong $CP$ problem is given by the 
Nelson-Barr mechanism based on spontaneous $CP$ violation \cite{nb}.
Since $CP$ invariance guarantees $\theta=0$ in this scenario, 
smallness of $\bar\theta$ can be explained if the spontaneous $CP$ violation 
occurs satisfying ${\rm arg}[\det({\cal M}_u{\cal M}_d)]=0$. 
A crucial problem is how simple models can be constructed so as 
to generate a $CP$ phase in the CKM matrix keeping $\bar\theta<10^{-10}$. 
For such an example among several models, one may consider a model proposed 
by Bento, Branco and Parada (BBP) \cite{bbp}, which is an extension of the SM 
with vector-like fermions and a complex singlet scalar. 
In this model, a $CP$ phase could appear in the CKM matrix when the $CP$ symmetry 
is spontaneously broken in the scalar sector \cite{scp}. 
It is caused via mixing between SM fermions and 
vector-like fermions mediated by the singlet scalar.\footnote{Extension of the 
model has been discussed from several phenomenological view points. 
For example, see \cite{ds-ext,llept,val}.} 
In their model, extra heavy vector-like down type quarks are introduced 
and $Z_2$ symmetry is imposed to control a down type quark mass matrix.
Unfortunately, one-loop corrections and contributions from higher dimension 
operators to the quark mass matrix could generate corrections, which could violate
$\bar\theta<10^{-10}$ \cite{dine}. 

In the lepton sector, long baseline neutrino oscillation experiments such as 
NOvA and T2K \cite{nova,t2k} suggest the existence of a $CP$ violating phase in the 
PMNS matrix \cite{pmns}. If lepton Yukawa couplings are assumed to be 
complex as in the quark sector, it can be also derived in the same way 
as the CKM matrix as long as neutrinos are massive.
A similar idea to the BBP model may be applicable to the 
lepton sector in order to explain the $CP$ phase in the PMNS matrix.
In that case, since a lepton mass matrix is irrelevant to the strong $CP$ problem,
no constraint on the mass matrix is imposed by it differently from the quark sector.
As a result, such an extension could be relevant to the recently confirmed muon 
anomalous magnetic moment which shows the deviation at 4.2 $\sigma$ 
from the SM prediction \cite{mug2}.
Several articles suggest that the existence of charged vector-like leptons 
could explain it \cite{veclep}.  
It seems to be an interesting issue whether this kind of framework 
could give any connection 
between the origin of a complex phase in the PMNS matrix and large deviation of 
the muon anomalous magnetic moment from the SM prediction. 

It is well known that baryon number asymmetry existing in the Universe \cite{basym}
cannot be understood in the SM. Leptogenesis based on out-of-equilibrium decay of
heavy right-handed neutrinos \cite{fy} is considered to be a 
most promising scenario for it. 
Although $CP$ asymmetry in their decay is a crucial parameter, it is difficult to fix
it and we have to treat it as a free parameter since the identification of relevant 
$CP$ phases is not easy. Additionally, since the mass of the lightest right-handed 
neutrino should be larger than $10^9$ GeV for successful leptogenesis \cite{di,ks} in usual 
scenarios for the small neutrino mass generation \cite{seesaw,scot}, 
reheating temperature is required to be higher than it.
It constrains possible inflation scenarios.

In this paper, we study these issues by considering a model based on the spontaneous 
$CP$ violation.
We show that the model can explain the $CP$ phases in the CKM and PMNS matrices
in a consistent way with the strong $CP$ problem.
We also clarify a $CP$ phase relevant to the $CP$ asymmetry
in the decay of the right-handed neutrinos and show that much lower reheating
temperature than $10^9$ GeV is allowed for successful leptogenesis in the model.

The remaining parts of the paper are organized as follows.
In section 2, we introduce our model and discuss its scalar sector.
We estimate reheating temperature expected in an inflation scenario supposed in the model. 
In section 3, we discuss a $CP$ phase in the CKM matrix and the strong 
$CP$ problem, and also the neutrino mass generation and $CP$ phases in the PMNS matrix.
We estimate $CP$ asymmetry in leptogenesis and show that leptogenesis occurs 
successfully at a low scale in a consistent way with the expected reheating temperature
in the supposed inflation scenario.  
We also show that the anomalous magnetic moment of muon suggested by 
the experiments cannot be explained in this model.
Section 4 is devoted to the summary of the paper. 

\section{Model for spontaneous $CP$ violation}
We consider an extension of the SM by introducing vector-like charged leptons $E_{L,R}$ 
and down type quarks $D_{L,R}$, right-handed neutrinos $N_j$, and several scalars, that is, 
a complex scalar $S$, a real scalar $\sigma$, and an inert doublet scalar $\eta$.  
We also impose a global discrete symmetry $Z_4\times Z_4^\prime$.
The model is intended to give a solution to the strong $CP$ problem and bring about 
$CP$ phases in the PMNS matrix simultaneously along the lines of the Nelson-Barr 
mechanism \cite{nb}. After spontaneous breaking of the discrete symmetry, 
the model is reduced to a scotogenic model for the neutrino mass \cite{scot} at low energy 
regions effectively. Representation of the introduced fields under 
$[SU(3)_C\times SU(2)_L\times U(1)_Y]\times Z_4 \times Z_4^\prime$ is 
summarized in Table 1. 
Since the SM contents are assumed to have no charge of $Z_4\times Z_4^\prime$, the invariant
Yukawa terms relevant to quarks are given as
\begin{eqnarray}
{\cal L}_q\supset \sum_{i=1}^3\left[\sum_{j=1}^3h_{ij}^d\bar q_{L_i}\tilde\phi d_{R_j}+ 
(y_i^dS+\tilde y_i^dS^\dagger)\bar D_L d_{R_i} \right]+ y_D\sigma\bar D_LD_R+ {\rm h.c.},  
\label{qlag} 
\end{eqnarray}
where $q_{L_i}$ and $d_{R_i}$ stand for the SM doublet and singlet 
quarks, respectively.\footnote{$\phi$ is an ordinary Higgs scalar. Definition 
$\tilde\phi=i\tau_2\phi^\ast$ is used.}  
Some new Yukawa terms are also introduced to charged leptons and neutrinos 
\begin{eqnarray}
{\cal L}_\ell&\supset& \sum_{i=1}^3\left[\sum_{j=1}^3h_{ij}^e
\bar\ell_{L_i}\tilde\phi e_{R_j}+ 
(y_i^eS+\tilde y_i^eS^\dagger)\bar E_L e_{R_i} 
+x_i\bar\ell_{L_i}\tilde \phi E_R\right] \nonumber \\
&+& (y_ES +\tilde y_E S^\dagger)\bar E_LE_R+ {\rm h.c.},
\label{llag} \\
{\cal L}_\nu&\supset&\sum_{j=1}^3\left[\sum_{i=1}^3 h_{ij}^\nu\bar\ell_{L_i}\eta N_j+ (y_{N_j}S+\tilde y_{N_j}S^\dagger)
\bar N_jN_j^c  + {\rm h.c.}\right], 
\label{nlag}
\end{eqnarray}
where $\ell_{L_i}$ and $e_{R_i}$ stand for the SM doublet and singlet leptons, respectively. 

\begin{figure}[t]
\begin{center}
\begin{tabular}{c|ccc||c|ccc}
      & SM & $Z_4$ & $Z_4^\prime$ &
&SM & $Z_4$ & $Z_4^\prime$ 
\\ \hline
$E_L$ & $({\bf 1},{\bf 1},-1)$  & 2 & 2 & 
$D_L$ & $({\bf 3},{\bf 1},-\frac{1}{3})$  & 2 & 2  \\
$E_R$ & $({\bf 1},{\bf 1},-1)$  & 0  & 0& 
$D_R$ & $({\bf 3},{\bf 1},-\frac{1}{3})$  & 0 & 2 \\
$N_j$ & $({\bf 1},{\bf 1},0)$  & 1 & 1 &
$S$ & $({\bf 1},{\bf 1},0)$  & 2  &  2 \\
$\eta$ & $({\bf 1},{\bf 2},-\frac{1}{2})$  & 3 & 3 &
$\sigma$ & $({\bf 1},{\bf 1},0)$ & 2 & 0    \\ \hline
\end{tabular}
\end{center}
\footnotesize{Table 1~~Representation of vector-like fermions and scalars  
added to the SM. In this table, 
SM stands for $SU(3)_C\times SU(2)_L\times U(1)_Y$. They play crucial roles 
in solving the strong $CP$ problem and also
in explaining $CP$ phases in the PMNS matrix, the neutrino mass, and dark 
matter. }
\end{figure}

Scalar potential invariant under the assumed symmetry can have a lot of terms.  
However, in the present study, we just assume rather restricted ones among them as
\begin{eqnarray}
V&=&V_1+V_2, \nonumber \\
V_1&=&\kappa_S(S^\dagger S)^2 
+\frac{1}{4}\kappa_\sigma\sigma^4+\frac{1}{2}\kappa_{S\sigma}(S^\dagger S)\sigma^2
+\kappa_{S\phi}(S^\dagger S)(\phi^\dagger\phi)  \nonumber \\
&+&\frac{1}{2}\kappa_{\sigma\phi}\sigma^2(\phi^\dagger\phi)
+m^2_S(S^\dagger S) +\frac{1}{2}m^2_\sigma\sigma^2 +V_b.  
\label{cppot} \\
V_2&=&\lambda_1(\phi^\dagger\phi)^2
+\lambda_2(\eta^\dagger\eta)^2+\lambda_3(\phi^\dagger\phi)(\eta^\dagger\eta)
+\lambda_4(\phi^\dagger\eta)(\eta^\dagger\phi) \nonumber \\ 
&+&\frac{\lambda_5}{2}\left[\frac{S}{M_\ast}(\eta^\dagger\phi)^2+{\rm h.c.}\right]
+m_\phi^2\phi^\dagger\phi+m_\eta^2\eta^\dagger\eta,
\label{npot}
\end{eqnarray}
where $M_\ast$ is a cut-off for physics relevant to the inert doublet $\eta$.
We list terms up to dimension 5. 
Several terms allowed under the imposed symmetry are assumed to be zero 
in this potential, for simplicity. Since $CP$ symmetry is assumed to be exact in the model, 
all the coupling constants in the Lagrangian are real. 

$V_b$ is composed of the $S$ number violating but $Z_4\times Z_4^\prime$ 
invariant terms such as $S^2$ and $S^4$ \cite{scp}.  
Spontaneous $CP$ violation could be caused in this part if $S$ gets  a vacuum 
expectation value (VEV) . 
As such a simple example of $V_b$, we consider
\begin{equation} 
V_b=\alpha(S^4 +S^{\dagger 4}) + \beta(S^2 +S^{\dagger 2} )\phi^\dagger\phi. 
\label{cpvpot}
\end{equation} 
If we express $S$ as $S=\frac{1}{\sqrt 2}\tilde Se^{i\rho}$, 
$\rho$ appears only in $V_b$ which can be rewritten as
\begin{equation}
V_b=\alpha\left(\tilde S^2\cos 2\rho+\frac{\beta}{4\alpha}\phi^\dagger\phi\right)^2
-\frac{\alpha}{2}\tilde S^4-\frac{\beta^2}{16\alpha}(\phi^\dagger\phi)^2.
\end{equation}
Thus, an angular component $\rho$ is fixed at this potential valley in the neutral field space.
It is expressed by using $\tilde S$ and a radial part $\phi_0$ of the neutral component  
of the doublet scalar $\phi$ as
\begin{equation}
\cos 2\rho=-\frac{\beta \phi_0^2}{4\alpha \tilde S^2},
\label{condcp}      
\end{equation} 
as long as the coupling constants $\alpha$ and $\beta$ take appropriate values.

Here, we specify the vacuum structure of this model.
We assume that these scalars take VEVs such as  
\begin{equation}
\langle S\rangle= \frac{u}{\sqrt 2}e^{i\rho_0}, \quad \langle \sigma\rangle= w,  \quad 
\langle\phi\rangle=\left(\begin{array}{c}\frac{v}{\sqrt 2} \\ 0 \end{array}\right),  \quad
 \langle\eta\rangle=0,
\end{equation} 
where $v(\equiv\langle\phi^0\rangle)=246~{\rm GeV}$ and $u,~w\gg v$ is 
assumed.\footnote{Although fine-tuning is required to realize this,
we do not discuss it further and just assume this hierarchical structure here.} 
Since $u \gg v$ is supposed, spontaneous $CP$ violation could occur and
$\rho_0\sim \frac{\pi}{4}$ is realized. 
Potential for the neutral scalars in $V_1$ at the potential valley defined by eq.~(\ref{condcp})
can be approximately expressed as
\begin{eqnarray}
V_1^0(S_R,S_I,\sigma)=
\frac{\kappa_\sigma}{4}(\sigma^2-w^2)^2+
\frac{\tilde\kappa_S}{4}(S_R^2+S_I^2-u^2)^2
+\frac{\kappa_{\sigma S}}{4}(\sigma^2-w^2)(S_R^2+S_I^2-u^2),
\label{potv}
\end{eqnarray}
where $|\kappa_{S\phi}|$ and $|\kappa_{\sigma\phi}|$ are assumed to be much 
smaller than others.
The coupling $\tilde \kappa_S$ is defined as
$\tilde\kappa_S=\kappa_S-2\alpha$.\footnote{We note that $\lambda_1$ is shifted to 
$\tilde\lambda_1=\lambda_1-\frac{\beta^2}{4\alpha}$ in $V_2$ due to an effect of $V_b$. } 
In order to guarantee the stability of the potential (\ref{potv})
these couplings should satisfy the conditions
\begin{equation}
\kappa_\sigma,~\tilde\kappa_S>0, \qquad 
4\tilde\kappa_S\kappa_\sigma>\kappa_{\sigma S}^2.
\label{stab}
\end{equation} 
Absolute values of these couplings could be constrained by a supposed 
inflation scenario as discussed later. 

It is useful to note that the imposed discrete symmetry $Z_4\times Z_4^\prime$ is
spontaneously broken to its diagonal subgroup $Z_2$ in this vacuum.
This $Z_2$ could stabilize the lightest field with its odd charge and guarantee the existence 
of candidates of dark matter (DM).  Since the remaining $Z_2$ keeps a uniqueness of 
the vacuum, the appearance of cosmologically dangerous stable domain walls 
associated to the breaking of discrete symmetry \cite{dw} is escapable.   

The neutral scalar sector characterizes the model depending on this vacuum.
A squared mass matrix for $\phi_0$, $S_R$, $S_I$ 
and $\sigma$ is given for a basis $\varphi^T=(\phi_0, S_R,S_I,\sigma)$ as
\begin{equation}
{\cal M}_s^2=\left(\begin{array}{cccc} 
2\tilde\lambda_1 v^2 & 
(\kappa_{S\phi}+2\beta)v u \cos\rho_0& 
(\kappa_{S\phi}-2\beta)v u\sin\rho_0
& \kappa_{\sigma\phi}v w\\
(\kappa_{S\phi}+2\beta)v u \cos\rho_0 &2(\tilde\kappa_S+4\alpha)u^2\cos^2\rho_0 & 
(\tilde\kappa_S-4\alpha)u^2\sin 2\rho_0 
& \kappa_{S\sigma}wu\cos\rho_0  \\    
(\kappa_{S\phi}-2\beta)v u\sin\rho_0&(\tilde\kappa_S-4\alpha)u^2\sin 2\rho_0   &
2(\tilde\kappa_S+4\alpha)u^2\sin^2\rho_0 
& \kappa_{S\sigma}wu\sin\rho_0 \\ 
\kappa_{\sigma\phi}v w&\kappa_{S\sigma}wu\cos\rho_0 
&\kappa_{S\sigma}wu\sin\rho_0 & 2\kappa_\sigma w^2
\end{array}\right).
\label{mscalar}
\end{equation}
If ${\cal M}_s^2$ is diagonalized as $O{\cal M}_s^2O^T={\cal M}^2_{s,{\rm diag}}$ by using an 
orthogonal matrix $O$, the mass eigenstate $\chi$ is related to $\varphi$ 
as $\chi=O\varphi$.
Since the couplings $|\kappa_{S\phi}|$ and 
$|\kappa_{\sigma\phi}|$ in eq.~(\ref{cppot}) are assumed to be sufficiently small 
and $v\ll u,w$ is satisfied, 
mixing of other scalars with $\phi_0$ is small enough not to affect the nature of 
the neutral Higgs scalar largely.  Moreover, we consider a case where
$|\kappa_{S\sigma}|u\ll\kappa_\sigma w$ is satisfied.
If we focus our study on such a case, 
$\chi_1\sim\phi_0$ and $\chi_4\sim\sigma$ are satisfied, and 
$\chi_2$ and $\chi_3$ are linear combinations of $S_R$ and $S_I$ as
\begin{equation}
\chi_2= S_R\cos\psi -S_I\sin\psi, \qquad \chi_3= S_R\sin\psi +S_I\cos\psi,
\label{eigens}
\end{equation}
where $\psi$ is found to be defined as
\begin{equation}
\tan 2\psi=-\frac{\tilde\kappa_S-4\alpha}{\tilde\kappa_S+4\alpha}\tan 2\rho_0.
\label{psi}
\end{equation}
If we suppose
$\rho_0\simeq\frac{\pi}{4}$, mass eigenvalues $m_i$ of these scalars $\chi_i$ are 
approximately evaluated as 
\begin{equation}
m_1^2\simeq 2\tilde\lambda_1 v^2, \quad m_2^2\simeq 2\tilde \kappa_S u^2, \quad
m_3^2 \simeq 8\alpha u^2, \quad
m_4^2\simeq 2\kappa_\sigma w^2.
\label{smass}
\end{equation}
Taking account of eqs.~(\ref{condcp}) and (\ref{psi}),
 $\rho_0$ and $\psi$ are found to be expressed as  
\begin{equation}
\rho_0\simeq\frac{\pi}{4}+\frac{\beta v^2}{8\alpha u^2}, \qquad
\psi\simeq\frac{\pi}{4}-\frac{\tilde\kappa+4\alpha}{\tilde\kappa-4\alpha}
\frac{\beta v^2}{8\alpha u^2}.
\label{rho}
\end{equation}

These singlet scalars could cause several effects on the phenomenology beyond the SM.
One of such issues is inflation of the Universe and reheating temperature expected from it.
Here, we consider $S_I$ as a candidate of inflaton.
Details of this inflation are discussed in Appendix A.
In this part, we only focus on reheating temperature realized in this inflation scenario 
through a perturbative process, which is expected to give a lower bound for
possible reheating temperature.

When the inflaton amplitude becomes $O(u)$ and the Hubble  parameter takes a value 
$H(u)=\left(\frac{\frac{1}{4}\tilde\kappa_Su^4}{3M_{\rm pl}^2}\right)^{1/2}$,
the inflaton is considered to start decaying through 
$S_I \rightarrow \phi^\dagger\phi$ in the case 
$y_j^d,  y_j^e, y_{N_j}>\sqrt{\tilde\kappa_S}$, for which the $S_I$ decay to fermions
$\bar D_LD_R, \bar E_LE_R$ and $N_{R_j}N_{R_j}$ are kinematically forbidden. 
Its decay width is estimated as
\begin{equation}
\Gamma\simeq \frac{1+\frac{1}{\sqrt 2}}{32\pi}\frac{\kappa_{S\phi}^2}{\sqrt{\tilde \kappa_S}}u,
\end{equation}
where $\alpha=0.1\kappa_S$ is assumed for simplicity.
If $\Gamma>H(u)$ is satisfied, instantaneous decay and thermalization are expected to 
occur. Then, reheating temperature is
determined by $\frac{\pi^2}{30}g_\ast T^4=\frac{1}{4}\tilde\kappa_S u^4$, where
$g_\ast$ represents relativistic degrees of freedom in the model.
We note that $\tilde\kappa_S$ is constrained by the CMB data as discussed in Appendix A. 
In the case $\Gamma<H(u)$, instantaneous decay cannot be applied, and 
reheating temperature should be estimated through 
$\Gamma=H(T)$ where 
$H(T)=\left(\frac{\frac{\pi^2}{30}g_\ast T^4}{3M_{\rm pl}^2}\right)^{1/2}$. 
Thus, the reheating temperature is fixed depending on the coupling constant $\kappa_{S\phi}$ as
\begin{equation}
T_R=\left\{\begin{array}{lr}
\displaystyle 8.7\times 10^3\left(\frac{\tilde\kappa_S}{10^{-6}}\right)^{1/4}
\left(\frac{u}{10^6~{\rm GeV}}\right)~{\rm GeV} &  \quad \displaystyle {\rm for}~~
|\kappa_{S\phi}|> C,  \\
\displaystyle 3.2\times 10^3\left(\frac{|\kappa_{S\phi}|}{10^{-9}}\right)
\left(\frac{10^{-6}}{\tilde\kappa_S}\right)^{1/4}
\left(\frac{u}{10^6~{\rm GeV}}\right)^{1/2}~{\rm GeV} & \quad \displaystyle {\rm for}~~
|\kappa_{S\phi}|< C, \\
\end{array}
\right.
\label{reheat}
\end{equation}
where $g_\ast=130$ is used and $C=2.7\times 10^{-9}
\left(\frac{\tilde\kappa_S}{10^{-6}}\right)^{1/2}\left(\frac{u}{10^6~{\rm GeV}}\right)^{1/2}$.
It suggests that this reheating temperature cannot be high 
enough for the thermal leptogenesis in the ordinary seesaw model 
for the neutrino mass \cite{di,ks}.
However, it is sufficiently high for successful leptogenesis in the present model.
We will see it later. 

Finally, we note here that the inflation scale $H_I$ is found to be much higher than the 
$CP$ breaking scale $u$ supposed in this model. It could bring about a 
serious domain wall problem 
caused by the spontaneous $CP$ violation \cite{wallcp}. 
However, since the inflation occurs through the inflaton which breaks 
the $CP$ symmetry, the $CP$ symmetry is violated during the inflation.
As a result,  the relevant domain wall is expected to be inflated away. 
It is not recovered throughout the inflaton oscillation period. 
Thus, the problem seems not to appear since the reheating temperature is 
lower than the $CP$ breaking scale $u$ as shown in eq.(\ref{reheat}).
It is noticeable that even such a low reheating temperature 
could make leptogenesis successful in the present model.

\section{Unified explanation of the $CP$ issues in the SM }
$CP$ issues in the SM could be treated in a unified way from the $CP$ phase caused 
by the spontaneous violation discussed in the previous section.
We discuss them in this section.
A $CP$ phase in the CKM matrix is shown to be derived using the Nelson-Barr mechanism.
The constraint on the $\bar\theta$ can be satisfied even if the radiative effects are taken 
into account. $CP$ phases in the PMNS matrix are also shown to be derived in the same way 
as the CKM phase. Baryon number asymmetry could be generated by 
thermal leptogenesis 
through the decay of right-handed neutrinos under the previously estimated reheating 
temperature. Sufficient $CP$ asymmetry in that decay 
is shown to be caused in a quantitatively fixed way.
 \subsection{CKM phase and solution for the strong $CP$ problem }
Yukawa interactions shown in ${\cal L}_q$ cause a $4\times 4$ 
mass matrix ${\cal M}_d^0$ for down-type quarks as
\begin{equation}
(\bar q_{L_i}, \bar D_L)\left(
\begin{array}{cc}
m_{ij}^d & 0 \\ {\cal F}_j^d & \mu_D \\
\end{array}\right)
\left(\begin{array}{c} d_{R_j} \\ D_R \\ \end{array} \right),
\label{qmass}
\end{equation}
where $m_{ij}^d=\frac{1}{\sqrt 2}h_{ij}^dv$, 
${\cal F}_j^d=\frac{1}{\sqrt 2}(y_j^d e^{i\rho_0} + \tilde y_j^d e^{-i\rho_0})u$, and $\mu_D=y_Dw$. 
We note that each component for $\bar q_{L_i}D_R$ in ${\cal M}_d^0$ is zero 
because of  the imposed discrete symmetry.
Since  an up-type quark mass matrix ${\cal M}_u$ is real by the assumed $CP$ invariance 
and ${\rm arg(det}{\cal M}_d^0)=0$ is fulfilled as found from eq.~(\ref{qmass}),
$\bar\theta=\theta+{\rm arg(det}{\cal M}_u{\cal M}_d^0)=0$ is still satisfied for 
$\rho_0\not=0$ after the spontaneous $CP$ violation.
This means that the strong $CP$ problem is solved at tree level by the Nelson-Barr 
mechanism.
On the other hand, a $CP$ phase in the CKM matrix could be caused from the 
$CP$ phase $\rho_0$. 

In order to see how the phase $\rho_0$ can generate the $CP$ phase in the CKM matrix,
we consider the diagonalization of a matrix ${\cal M}_d^0{\cal M}_d^{0^\dagger}$ 
by a $4\times 4$ unitary matrix $V_L$ as $V_L{\cal M}_d^0{\cal M}_d^{0\dagger} V_L^\dagger$.
It may be expressed as 
\begin{equation}
\left(\begin{array}{cc} A & B \\ C& D \\\end{array}\right)
\left(\begin{array}{cc} m^dm^{d\dagger} & 
m^d{\cal F}^{d\dagger} \\ 
  {\cal F}^dm^{d\dagger} & \mu_D^2 +{\cal F}^d{\cal F}^{d\dagger} \\
\end{array}\right)
\left(\begin{array}{cc} A^\dagger & C^\dagger \\ B^\dagger 
& D^\dagger \\\end{array}\right)=
\left(\begin{array}{cc}\tilde m_d^2 & 0 \\ 0 &\tilde M_D^2 \\\end{array}\right),
\label{dmass}
\end{equation}
where a $3\times 3$ matrix $\tilde m_d^2$ in the right-hand side is diagonal 
in which the generation indices are abbreviated. Eq.~(\ref{dmass}) requires
\begin{eqnarray}
   && m^dm^{d\dagger}=A^\dagger\tilde m_d^{2}A+ C^\dagger \tilde M_D^2C, \qquad
  {\cal F}^dm^{d\dagger}=B^\dagger \tilde m_d^2A+ D^\dagger \tilde M^2_DC, \nonumber \\
   && \mu_D^2+{\cal F}^d{\cal F}^{d\dagger}=
B^\dagger \tilde m_d^2 B+ D^\dagger \tilde M_D^2 D.
\end{eqnarray}  
Since $\mu_D^2 +{\cal F}^d{\cal F}^{d\dagger}$ could be much larger than each 
component of ${\cal F}^dm^{d\dagger}$, we find that $B, C$ and $D$ can be approximated as
\begin{equation}
  B\simeq -\frac{Am^d{\cal F}^{d\dagger}}{\mu_D^2+{\cal F}^d{\cal F}^{d\dagger}},
 \qquad C\simeq\frac{{\cal F}^d m^{d\dagger}}{\mu_D^2+{\cal F}^d{\cal F}^{d\dagger}},
   \qquad D\simeq 1,
\end{equation}
which guarantee the approximate unitarity of the matrix $A$. 
In such a case, it is also easy to find that
\begin{equation}
A^{-1}\tilde m_d^2A= m^dm^{d\dagger} -\frac{1}{\mu_D^2+{\cal F}^d{\cal F}^{d\dagger}}
m^d{\cal F}^{d\dagger}{\cal F}^dm^{d\dagger}.
\label{mix}
\end{equation}
The right-hand side is an effective mass matrix of the 
light down-type quarks which is derived through the mixing with the extra heavy quarks. 
Since the second term can have complex phases in off-diagonal 
components unless $\tilde y_j^d$ is equal to $y_j^d$, the matrix $A$ could be complex. 
Complex phases in the matrix $A$ could have a substantial magnitude
since the second term is comparable with the first term as long as 
$\mu_D^2<{\cal F}^d{\cal F}^{d\dagger}$ is satisfied.

As a reference, we show an example of the CKM matrix obtained in this scenario 
by assuming that the up-type quark mass matrix is diagonal. In this case, the CKM matrix 
is given as $V_{\rm CKM}=A$. If we take the relevant VEVs as
\begin{equation}
 u=10^6~{\rm GeV} ,  \qquad  w=10^5~{\rm GeV},  
\label{vevs}
\end{equation}
and Yukawa coupling constants as
\begin{eqnarray}
&&y^d=(0,5.2\times 10^{-4},0), \quad  \tilde y^d=(0,0,1.2\times 10^{-3}), \quad  y_D= 10^{-2}, \nonumber \\
&&h_{11}^d=6.0\times 10^{-6}, \quad  h_{22}^d=6.5\times 10^{-4}, \quad h_{33}^d=3.5\times 10^{-2}, 
\nonumber \\
&& h_{12}^d=h_{21}^d=1.45\times 10^{-4},\quad
h_{13}^d=h_{31}^d=7.0\times 10^{-5}, \quad h_{23}^d=h_{32}^d=1.6\times 10^{-3},
\label{qyuk}
\end{eqnarray}
the mass eigenvalues of the down-type quarks are obtained as
\begin{equation}
\tilde m_{d_1}=4.7~{\rm MeV},   \quad 
\tilde m_{d_2}= 95~{\rm MeV}, \quad \tilde m_{d_3}=4.2~{\rm GeV} , \quad  
\tilde M_D=1646~{\rm GeV}.
\end{equation} 
The CKM matrix and the Jarlskog invariant $J_q$ \cite{jrl} are determined as
\begin{equation}
V_{CKM}=\left(\begin{array} {ccc}
0.974 & 0.225 & 0.008 \\ 0.225 & 0.973 & 0.047 \\ 0.003 & 0.048 & 0.999 \\
\end{array} \right), \qquad 
J_q=1.64\times 10^{-5},
\end{equation}
where the absolute values for the components of $V_{CKM}$ are presented.
This example suggests that suitable parameters could reproduce the experimental results
well in this framework. 

\begin{figure}[t]
\begin{center}
\includegraphics[width=14cm]{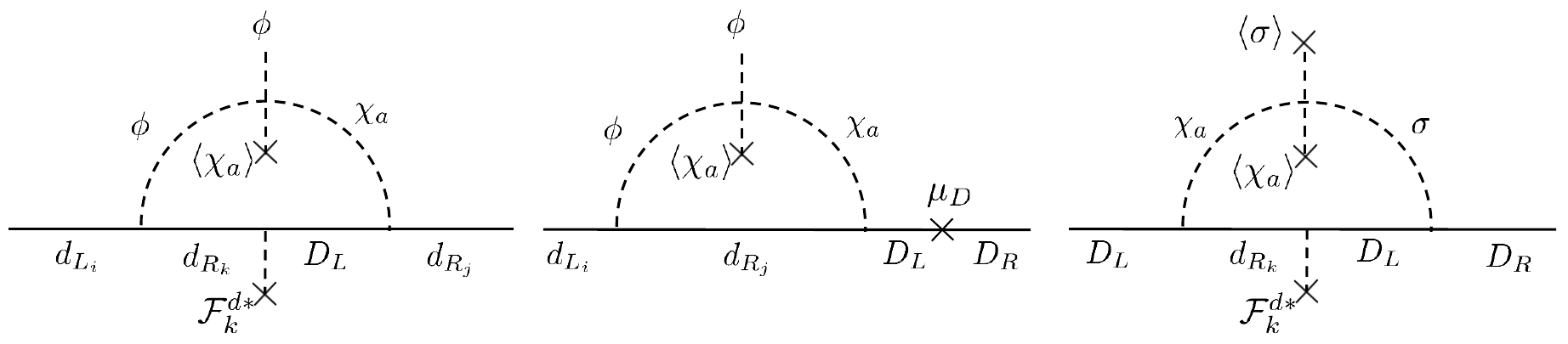}
\vspace*{-3mm}
\end{center}
\footnotesize{Fig. 1~~One-loop diagrams which give complex contributions to 
the down-type quark mass matrix ${\cal M}_d$. 
Each diagram corresponds to $\delta h^d_{ij}$, $\delta f^d_i$ and $\delta\mu_D$ 
from left to right, respectively.}
\end{figure}

For the strong $CP$ problem, eq.~(\ref{qmass}) does not mean to give a stable solution. 
One-loop radiative effects and higher order effective operators could give complex 
corrections to the Yukawa couplings \cite{dine}, which add $CP$ violating contributions to 
each component of the mass matrix (\ref{qmass}). 
Since they could violate the constraint $\bar\theta<10^{-10}$ easily,  we need to examine
whether the corrections are small enough to give a satisfactory solution for the 
strong $CP$ problem. 
One-loop complex corrections to the coupling constant $h^d_{ij}$,  a coupling constant
$f_i^d$ for the operator $\bar d_{L_i}\tilde\phi D_R$ which is zero at the tree level, 
and the mass $\mu_D$ are 
caused by diagrams shown in Fig.~1, respectively.
If we note that relevant Yukawa interactions in eq.~(\ref{qlag}) can be 
rewritten by using eq.~(\ref{eigens}) as 
\begin{eqnarray}
&&\sum_{j=1}^3\Big[h^d_{ij}\tilde\phi\bar d_{L_j}d_R+
\frac{1}{\sqrt 2}\left\{(y_j^d+\tilde y_j^d)\cos\psi-
i(y_j^d-\tilde y_j^d)\sin\psi\right\}\chi_2\bar D_L d_{R_j}  \nonumber\\
&&+\frac{1}{\sqrt 2}\left\{(y_j^d+\tilde y_j^d)\sin\psi+
i(y_j^d-\tilde y_j^d)\cos\psi\right\}\chi_3
\bar D_L d_{R_j}+{\rm h.c.}\Big],
\end{eqnarray}
we find that they can be estimated, respectively, as
\begin{eqnarray}
\delta h^d_{ij}&\simeq&\frac{1}{32\pi^2}
\ln\left(\frac{v^2}{u^2}\right)\sum_{k=1}^3h_{ik}^d\left\{(y^d_k+\tilde y_k^d)\cos\rho_0
-i(y_k^d-\tilde y_k^d)\sin\rho_0\right\}   \nonumber \\
&\times&\Big[\frac{\kappa_{S\phi} u^2}{m_2^2}
\left\{(y_j^d+\tilde y_j^d)\cos\psi-i(y_j^d-\tilde y_j^d)\sin\psi\right\}
\cos(\rho_0+\psi)  \nonumber \\
&&+\frac{\kappa_{S\phi} u^2}{m_3^2}\left\{(y_j^d+\tilde y_j^d)\sin\psi
+i(y_j^d-\tilde y_j^d)\cos\psi\right\}\sin(\rho_0+\psi) \Big],  \nonumber \\
\delta f^d_i&\simeq&\frac{\sqrt 2}{32\pi^2}
\ln\left(\frac{v^2}{u^2}\right)\sum_{k=1}^3h_{ik}^d \nonumber \\
&\times&\Big[\frac{\kappa_{S\phi}u\mu_D}{m^2_2}
\left\{(y^d_k+\tilde y_k^d)\cos\psi-i(y_k^d-\tilde y_k^d)\sin\psi\right\}
\cos(\rho_0+\psi)  \nonumber \\
&&+\frac{\kappa_{S\phi}u\mu_D}{m^2_3}
 \left\{(y^d_k+\tilde y_k^d)\sin\psi+i(y_k^d-\tilde y_k^d)\cos\psi\right\}
\sin(\rho_0+\psi)\Big],  \nonumber \\
\delta\mu_D&\simeq&\frac{\mu_D}{32\pi^2}
\sum_{k=1}^3\left\{(y^d_k+\tilde y_k^d)\cos\rho_0
-i(y_k^d-\tilde y_k^d)\sin\rho_0\right\}   \nonumber \\
&\times&\Big[\frac{\kappa_{S\sigma} u^2}{m_4^2-m_2^2}\ln\left(\frac{m_4^2}{m_2^2}\right)
\left\{(y_j^d+\tilde y_j^d)\cos\psi-i(y_j^d-\tilde y_j^d)\sin\psi\right\}
\cos(\rho_0+\psi)  \nonumber \\
&&+\frac{\kappa_{S\sigma} u^2}{m_4^2-m_3^2}\ln\left(\frac{m_4^2}{m_3^2}\right)
\left\{(y_j^d+\tilde y_j^d)\sin\psi+i(y_j^d-\tilde y_j^d)\cos\psi\right\}
\sin(\rho_0+\psi) \Big],  
\label{onel}
\end{eqnarray}
where $m_2^2$, $m_3^2$ and $m_4^2$ are the scalar mass eigenvalues given in eq.~(\ref{smass}).

On the other hand, higher order operators which give complex contribution to them 
at low energy regions come from dimension-6 ones
\begin{eqnarray}
\frac{S^2}{M_{\rm pl}^2}\bar d_L\tilde\phi d_R,  \qquad 
\frac{\sigma S}{M_{\rm pl}^2}\bar d_L\tilde\phi D_R, 
\qquad \frac{S^2}{M_{\rm pl}^2}\sigma \bar D_LD_R,
\label{ho}
\end{eqnarray}
where the $O(1)$ coupling constants are supposed for them.
Since the dominant contributions are expected to come from the one-loop contributions 
in the case $\frac{u}{M_{\rm pl}}=O(10^{-12})$, the mass matrix of the down type 
quarks is modified to
\begin{equation}
{\cal M}_d={\cal M}_d^0\left[ {\bf 1} +({\cal M}_d^0)^{-1}
\left(\begin{array}{cc}\delta h^dv  & \delta f^d v \\ \delta{\cal F}^d & \delta\mu_D\\
\end{array}\right)\right].
\end{equation}
Since the second term is much smaller than the first term in the right-hand side, 
 $\bar\theta={\rm arg(det}{\cal M}_d)$ 
can be estimated as
\begin{eqnarray}
\bar\theta&=& {\rm Im}\left[{\rm tr}\left\{({\cal M}_d^0)^{-1}
\left(\begin{array}{cc}\delta h^dv  & \delta f^d v \\ \delta{\cal F}^d & \delta\mu_D\\
\end{array}\right)\right\}\right]  \nonumber \\
&=&{\rm Im}\left[{\rm tr}((h^d)^{-1}\delta h^d)-
\frac{1}{\mu_D}\left({\cal F}^d (h^d)^{-1}\delta  f^d-\delta\mu_D\right)
\right] \nonumber \\
&=&\frac{1}{128\pi^2}
\frac{\kappa_{S\sigma}u^2}{\kappa_\sigma w^2}
\ln\left(\frac{\tilde\kappa_S}{4\alpha}\right)\sin 2(\rho_0+\psi)
\sum_{j=1}^3(y_j^{d2}- \tilde y_j^{d2}),
\end{eqnarray}
where we use eq.~(\ref{onel}) in the last equality. It is caused by 
${\rm Im}[\frac{\delta\mu_D}{\mu_D}]$
as a result of cancellation between other contributions.

If we use eq.~(\ref{rho}) for $\rho_0$ and $\psi$ and the parameters 
given in eq.~(\ref{qyuk}), which fixes 
$\sum_j(y_j^{d2}-\tilde y_j^{d2})$ to $O(10^{-6})$, 
the constraint $|\bar\theta|<10^{-10}$ can be expressed as
\begin{equation}
\frac{|\kappa_{S\sigma}|}{\kappa_\sigma}\frac{v^2}{w^2}
\ln\left(\frac{\tilde\kappa_S}{4\alpha}\right) < 10^{-2}.
\end{equation}
This condition can be easily satisfied for the supposed couplings by taking account of 
$\frac{v^2}{w^2}=O(10^{-6})$.  
In relation to this, 
it may be useful to note that dominant one-loop correction to $\kappa_{S\sigma}$ 
caused by the fermion loop could be estimated as
\begin{equation}
\delta\kappa_{S\sigma}=\frac{1}{16\pi^2}\sum_{k=1}^3y_D^2(y_k^{d2}+\tilde y_k^{d2})
\ln\frac{M_{\rm pl}^2}{u^2}.
\end{equation}
It is clear that this correction does not contradict the above condition.
The present analysis shows that the strong $CP$ problem can be solved in the model
even if the radiative effects are taken into account.
Here, on the points suggested in \cite{dine} we should note that the above result 
is obtained under the assumption that the couplings $\kappa_{S\phi}$ and 
$\kappa_{\sigma\phi}$ of the new singlet scalars with the Higgs scalar are 
sufficiently small, and additional fine-tuning is required in the scalar sector. 
In this sense, we might consider that the strong $CP$ problem is replaced with the small 
Higgs mass problem in this model.

\subsection{$CP$ phases in the PMNS matrix and DM}
A $CP$ phase can appear in the PMNS matrix through the 
couplings of the singlet $S$ with the vector-like charged leptons
in the same way as the CKM matrix case. 
In fact, the Yukawa interactions shown in ${\cal L}_\ell$ cause a $4\times 4$ 
mass matrix ${\cal M}_e$ as
\begin{equation}
(\bar\ell_{L_i}, \bar E_L)\left(
\begin{array}{cc}
m^e_{ij} & {\cal G}_i \\ {\cal F}^e_j & \mu_E \\
\end{array}\right)
\left(\begin{array}{c} e_{R_j} \\ E_R \\ \end{array} \right),
\label{lmass}
\end{equation}
where $m^e_{ij}=h_{ij}^ev$, 
${\cal F}_j^e=\frac{1}{\sqrt 2}(y_j^e e^{i\rho_0} + \tilde y_j^e e^{-i\rho_0})u$, 
${\cal G}_i=\frac{1}{\sqrt 2}x_i v$ and $\mu_E=\frac{1}{\sqrt 2}
(y_Ee^{i\rho_0}+\tilde y_Ee^{-i\rho_0})u$.
The difference from ${\cal M}^0_d$ appears in nonzero components ${\cal G}_i$ 
and the mass $\mu_E$.  Following the CKM case, we consider the diagonalization of a matrix 
${\cal M}_e{\cal M}_e^\dagger$ by a $4\times 4$ unitary matrix 
$\tilde V_L$ as $\tilde V_L{\cal M}_\ell{\cal M}_\ell^\dagger \tilde V_L^\dagger$.
It can be represented as 
\begin{equation}
\left(\begin{array}{cc} \tilde A & \tilde B \\ \tilde C& \tilde D \\\end{array}\right)
\left(\begin{array}{cc} m^em^{e\dagger}+{\cal G}{\cal G}^\dagger & 
m^e{\cal F}^{e\dagger}+\mu_E^\ast{\cal G} \\ 
  {\cal F}^em^{e\dagger}+{\cal G}^\dagger \mu_E & |\mu_E|^2 +{\cal F}^e{\cal F}^{e\dagger} \\
\end{array}\right)
\left(\begin{array}{cc} \tilde A^\dagger & \tilde C^\dagger \\ \tilde B^\dagger 
& \tilde D^\dagger \\\end{array}\right)=
\left(\begin{array}{cc}\tilde m_e^{2} & 0 \\ 0 &\tilde M_E^2 \\\end{array}\right),
\label{mass}
\end{equation}
where a $3\times 3$ matrix $\tilde m_e^2$ in the right-hand side is diagonal again. 
Eq.~(\ref{mass}) requires
\begin{eqnarray}
   && m^em^{e\dagger}+{\cal G}{\cal G}^\dagger=\tilde A^\dagger\tilde m_e^2\tilde A+
\tilde C^\dagger \tilde M_E^2\tilde C, \qquad
  {\cal F}^em^{e\dagger}+{\cal G}^\dagger\mu_E=\tilde B^\dagger \tilde m_e^2\tilde A+ 
\tilde D^\dagger \tilde M^2_E\tilde C, \nonumber \\
   && |\mu_E|^2+{\cal F}^e{\cal F}^{e\dagger}=
\tilde B^\dagger \tilde m_e^2\tilde B+ \tilde D^\dagger \tilde M_E^2\tilde D.
\end{eqnarray}  
Since $|\mu_E|^2 +{\cal F}^e{\cal F}^{e\dagger}$ is much larger than each 
components of ${\cal F}^em^{e\dagger}+{\cal G}\mu_E^\ast$,
we find that $\tilde B, \tilde C$ and $\tilde D$ can be approximately expressed 
 in the same way as the case of the CKM matrix,
\begin{equation}
  \tilde B\simeq -\frac{\tilde A(m^e{\cal F}^{e\dagger}
+\mu_E^\ast{\cal G})}{|\mu_E|^2+{\cal F}^e{\cal F}^{e\dagger}},
 \qquad \tilde C\simeq\frac{{\cal F}^e m^{e\dagger}+{\cal G}^\dagger \mu_E}{|\mu_E|^2
+{\cal F}^e{\cal F}^{e\dagger}},
   \qquad \tilde D\simeq 1.
\end{equation}
These again guarantee the approximate unitarity of the matrix $\tilde A$. 
In such a case, it is also easy to find the relation
\begin{equation}
\tilde A^{-1}\tilde m^{e2}\tilde A= m^em^{e\dagger}+{\cal G}{\cal G}^\dagger 
-\frac{1}{|\mu_E|^2+{\cal F}^e{\cal F}^{e\dagger}}
(m^e{\cal F}^{e\dagger}+\mu_E^\ast{\cal G})({\cal F}^em^{e\dagger}+\mu_E{\cal G}^\dagger ).
\label{mix}
\end{equation}
The charged lepton effective mass matrix $\tilde m_e$ is obtained as a result 
of the mixing between the light charged leptons and the extra heavy leptons. 
If $\tilde y_j^e$ is not equal to $y_j^e$ and $|\mu_E|^2<{\cal F}^e{\cal F}^{e\dagger}$, 
the matrix $\tilde A$ could have a large $CP$ phase.

The mass of neutrinos can be generated through the radiative effect 
as in the scotogenic model since the present model is reduced to it effectively after $S$ 
gets the VEV. 
As found in eq.~(\ref{nlag}), $N_j$ has Yukawa couplings with $\nu_{L_i}$ and $\eta$. 
However, since $\eta$ is assumed to have no VEV,
neutrino masses are not generated at tree level but generated at one-loop level.
We note that a small complex effective coupling constant 
$\tilde\lambda_5=\lambda_5\frac{u}{M_\ast}e^{i\rho_0}$ is induced even 
in the case $\lambda_5=O(1)$. 
The effective coupling $\frac{\tilde\lambda_5}{2}(\eta^\dagger\phi)^2+{\rm h.c.}$ 
brings about a small mass difference between the real and imaginary components of $\eta^0$. 
As its result, the one-loop diagram with $N_j$ and $\eta^0$ in internal lines gives
a nonzero contribution to the neutrino mass.  If we note that the mass of $N_j$ are generated 
through the coupling  $(y_{N_j}S+\tilde y_{N_j}S^\dagger)\bar N_jN_j^c$ in eq.~(\ref{nlag}), 
the neutrino mass is found to be expressed as
\begin{eqnarray}
&&{\cal M}_{\nu_{ij}}=\sum_{k=1}^3
 h_{ik}^\nu h_{jk}^\nu\Lambda_k e^{i(\theta_k+\rho_0)}, \nonumber \\
&&\Lambda_k=\frac{
|\tilde\lambda_5|\langle\phi\rangle^2}{8 \pi^2M_{N_k}}
  \left[\frac{M_{N_k}^2}{M_\eta^2-M_{N_k}^2}
    \left(1+\frac{M_{N_k}^2}{M_\eta^2-M_{N_k}^2}
    \ln\frac{M_{N_k}^2}{M_\eta^2}\right) \right],
\label{lnmass}
\end{eqnarray}
where $M_{N_k}$, $\theta_k$ and $M_\eta^2$ are defined as
\begin{eqnarray}
&&M_{N_k}=(y_{N_k}^2+\tilde y_{N_k}^2+2y_{N_k}\tilde y_{N_k}\cos 2\rho_0)^{1/2}u, \quad
\tan\theta_k=\frac{y_{N_k}-\tilde y_{N_k}}{.y_{N_k}+\tilde y_{N_k}}\tan\rho_0, \nonumber\\
&&M_\eta^2=m_\eta^2+(\lambda_3+\lambda_4)\langle\phi\rangle^2.
\label{theta}
\end{eqnarray}
The formula (\ref{lnmass}) can explain small neutrino masses required by the neutrino oscillation 
data \cite{pdg} even for $N_j$ with the mass of order TeV scale since the smallness 
of $|\tilde\lambda_5|$ is naturally guaranteed by $u\ll M_\ast$ as addressed above.

If we consider that the matrix ${\cal M}_\nu$ is diagonalized by a unitary matrix
$U_\nu$ such as $U_\nu^T{\cal M}_\nu U_\nu={\cal M}_\nu^{\rm diag}$, the PMNS matrix is obtained
as $V_{PMNS}=\tilde A^\dagger U_\nu$ where $\tilde A$ is fixed through eq.~(\ref{mix}).
Since the matrix $\tilde A$ is expected to be almost diagonal from
hierarchical charged lepton masses, the structure of $V_{PMNS}$ 
is considered to be mainly determined by $U_\nu$ in the neutrino sector.
It is well known that tribimaximal mixing cannot realize a nonzero mixing 
angle $\theta_{13}$, which is required by the neutrino oscillation data. 
However, if the matrix $\tilde A$ can compensate
this fault, a desirable $V_{PMNS}$ may be derived as $V_{PMNS}=\tilde A^\dagger U_\nu$
even if $U_\nu$ takes the tribimaximal form.
The tribimaximal structure in the neutrino sector can be easily realized 
if we adopt a simple assumption for neutrino Yukawa couplings such as \cite{tribi}
\begin{eqnarray}
h^\nu_{1j}=0, \quad h^\nu_{2j}=h^\nu_{3j}=h_j~~(j=1,2), \qquad 
h^\nu_{13}=h^\nu_{23}=-h^\nu_{33}=h_3.
\label{ny}
\end{eqnarray}
Under this assumption, the mass eigenvalues of ${\cal M}_\nu$ given in eq.~(\ref{lnmass}) are fixed as
\begin{equation}
m_1^\nu=0, \quad m_2^\nu=3h_3^2\Lambda_3, \quad 
m_3^\nu=2[h_1^4\Lambda_1^2+h_2^4\Lambda_2^2+
2h_1^2h_2^2\Lambda_1\Lambda_2\cos(\theta_1-\theta_2)]^{1/2}.
\label{nmass}
\end{equation}
This suggests that the squared mass differences required  by the neutrino oscillation 
data can be realized if both $h_2$ and $h_3$ take values of $O(10^{-2})$ for 
$\Lambda_{2,3}=O(1)$~eV, which can be realized for TeV scale $M_\eta$ and $M_{N_j}$.
 The diagonalization matrix $U_\nu$ can be expressed as
\begin{equation}
U_\nu=\left(\begin{array}{ccc}\frac{2}{\sqrt 6} & \frac{1}{\sqrt 3} & 0\\
\frac{-1}{\sqrt 6} & \frac{1}{\sqrt 3} & \frac{1}{\sqrt 2}\\
\frac{1}{\sqrt 6} & \frac{-1}{\sqrt 3} & \frac{1}{\sqrt 2}\\ \end{array}\right)
\left(\begin{array}{ccc}1 & 0 & 0\\
0 & e^{-i\gamma_1} & 0\\
0 & 0 &e^{-i\gamma_2} \\ \end{array}\right),  
\end{equation}
where $\gamma_{1}$ and $\gamma_{2}$ are defined as
\begin{equation}
\gamma_1=\frac{\theta_3}{2}, \quad
\gamma_2=\frac{1}{2}\tan^{-1}\left[\frac{h_1^2\Lambda_1\sin\theta_1+h_2^2\Lambda_2\sin\theta_2}
{h_1^2\Lambda_1\cos\theta_1+h_2^2\Lambda_2\cos\theta_2}\right].
\end{equation}

We examine whether the present scenario works in this simple tribimaximal case by
fixing the relevant parameters.
For this purpose, we use the values of $u$ and $w$ given in eq.~(\ref{vevs}).
Other input parameters are taken to be
\begin{eqnarray}
&&y^e=(0,10^{-4},0), ~ \tilde y^e=(0,0,3.3\times 10^{-5}), ~ x=(2.2\times 10^{-4},1.5\times 10^{-3},
8\times 10^{-3}) , 
\nonumber\\
&& y_E=\tilde y_E= 3.3\times 10^{-6}, ~h_{11}^e=5.7\times 10^{-6}, ~ h_{22}^e=1.2\times 10^{-4}, ~
 h_{33}^e=7\times 10^{-3}, 
\nonumber \\
&& h_{12}^e=h_{21}^e=4\times 10^{-5},~
h_{13}^e=h_{31}^e=1.7\times 10^{-6}, ~ h_{23}^e=h_{32}^e=4.7\times 10^{-4} .
\end{eqnarray}
These give mass eigenvalues of the charged leptons as
\begin{equation}
\tilde m_{e_1}=0.59~{\rm MeV},   \quad 
\tilde m_{e_2}= 0.106~{\rm MeV}, \quad \tilde m_{e_3}=1.81~{\rm GeV} , \quad  
\tilde M_E=3165~{\rm GeV}.
\end{equation} 
The PMNS matrix and the Jarlskog invariant $J_\ell$ are determined 
as\footnote{Here, we note that $J_\ell$ does not depend on the Majorana phases.}
\begin{equation}
V_{PMNS}=\left(\begin{array} {ccc}
0.837 & 0.526 & 0.149 \\ 0.412 & 0.672 & 0.615 \\ 0.360 & 0.521 & 0.774 \\
\end{array} \right), \qquad 
J_\ell=-0.032,
\end{equation}
where the absolute values are presented for each element of $V_{PMNS}$.
We find that these are a rather good realization of the experimental results. 
 
The imposed global symmetry in the model could guarantee the stability of 
some neutral fields and present candidates of DM. 
The present model has an inert doublet scalar $\eta$ and three right-handed 
neutrinos $N_j$ which are the only fields with the odd parity of the remnant $Z_2$ symmetry.
Since $\eta$ is assumed to have no VEV, $Z_2$ remains as an exact symmetry.
It guarantees the stability of the lightest one with its odd parity as in the ordinary 
scotogenic model where DM candidates are included in the model as crucial ingredients. 
Possible DM candidates are the lightest $N_j$ or the lightest neutral component of $\eta$.
Both of them can have TeV scale mass in a consistent way with the neutrino oscillation data.
In the case where $N_1$ is DM with a TeV scale mass, the Yukawa coupling 
$h_{i1}^\nu$ should be large to decrease its relic density to the required amount. 
It causes a dangerous lepton flavor violating process such as 
$\mu\rightarrow e\gamma$ \cite{kms}.
On the other hand, the lightest neutral component of $\eta$ can be a good DM candidate
without causing serious phenomenological contradiction.
It has been extensively studied as a CDM candidate, and it has been found that its thermal 
relics in this mass range could have a suitable amount if the quartic couplings 
$\lambda_3$ and $\lambda_4$ in eq.~(\ref{npot}) take suitable values \cite{etadm,ks}.

\subsection{$CP$ asymmetry in leptogenesis}
In the ordinary scotogenic model for the neutrino mass generation, 
required baryon number asymmetry cannot be generated 
through thermal leptogenesis due to the decay of the lightest right-handed neutrino $N_1$  
unless its mass is larger than $O(10^8)$ GeV \cite{ks}. 
For sufficient production of the thermal abundance of $N_1$, 
large neutrino Yukawa couplings $h_{i 1}^\nu$ are required and then larger $N_1$ 
mass is needed to make neutrino masses suitable for the explanation of the neutrino 
oscillation data.
On the other hand, small couplings $h_{i 1}^\nu$ are favored to sufficiently suppress
the washout of lepton number asymmetry generated through the $N_1$ decay.
These fix the above mentioned lower bound of the $N_1$ mass and also the lower 
bound of the reheating temperature.

Fortunately, this bound could be relaxed automatically in the present model.
$N_1$ could be generated in the thermal bath through other built-in processes, that is,
the scattering of the vector-like fermions such as $\bar E_LE_R \rightarrow N_jN_j$,
$\bar E_Le_{R_i} \rightarrow N_jN_j$, and $\bar D_Ld_{R_j} \rightarrow N_jN_j$, 
which are mediated by the neutral scalars $S_R$ and $S_I$.
The second and third ones among these are expected to give dominant contributions 
since relevant Yukawa coupling constants take larger values in the previous examples. 
For example, the reaction rate of the second process can be roughly estimated at the temperature
 $T~(>\tilde M_E)$ as
\begin{eqnarray}
\Gamma^{(Ee)}_S(ij)&\simeq& \frac{T^5}{64\pi}\Big[(y_i^{e2}+\tilde y_i^{e2})(y_{N_j}^2+\tilde y_{N_j}^2)
\left(\frac{1}{m_2^4}+\frac{1}{m_3^4}\right)\nonumber \\
 &+&2\left\{(y_i^{e}\tilde y_{N_j}+\tilde y_i^{e}y_{N_j})^2-(y_i^ey_{N_j}-\tilde y^e_i\tilde y_{N_j})^2\right\}
\frac{1}{m_2^2m_3^2}\Big],
\end{eqnarray}
where $m_2$ and $m_3$ are given in eq.~(\ref{smass}).
Since this process is irrelevant to the neutrino Yukawa couplings $h_{i 1}^\nu$,
they can take sufficiently small values so as to make the washout process 
ineffective.\footnote{This is allowed since the squared mass 
differences required to explain the neutrino oscillation data can be caused by two 
right-handed neutrinos $N_2$ and $N_3$ only.}
The heavy lepton $E$ is expected to be in the thermal equilibrium through 
the SM gauge interactions if reheating temperature $T_R$ and its mass $\tilde M_E$ 
satisfy $\tilde M_E<T_R$.
Thus, if the reaction rate $\Gamma^{(Ee)}_S(i1)$ of this scattering and the Hubble parameter $H$
satisfy a condition $\Gamma^{(Ee)}_S(i1)\sim H(T)$ at the temperature $T$, 
$N_1$ could be produced sufficiently as long as the temperature $T$ is larger than $M_{N_1}$. 
In fact, if we apply the parameters used in the previous example, this condition is found 
to be satisfied around the temperature\footnote{A value of $\tilde\kappa_S$ is referred to
the result given in eq.~(\ref{kap}).}
\begin{equation} 
T\sim 2.3\times 10^3\left(\frac{10^{-4}}{y^e_i}\right)^{2/3}
\left(\frac{10^{-3}}{y_{N_1}}\right)^{2/3}\left(\frac{\tilde\kappa_S}{10^{-6}}\right)^{2/3}
\left(\frac{u}{10^6{\rm GeV}}\right)^{4/3}~{\rm GeV}.
\end{equation}
The estimated lower bound of the reheating temperature in eq.~(\ref{reheat}) could be higher 
than this. If $M_{N_1}$ takes a value of $O(1)$~TeV, 
its number density is expected to reach the relativistic equilibrium value 
$n_{N_1}^{\rm eq}(T)$ of $O(10^{-3})$. 

On the $CP$ asymmetry $\varepsilon$ of the $N_1$ decay,
if we note that all the Yukawa couplings $h_{ij}^\nu$ are real and 
it is independent of the PMNS matrix, $\varepsilon$ is found to be expressed as
\begin{equation}
\varepsilon=\frac{1}{8\pi}\sum_{j=2,3}
\frac{(\sum_i h_{i1}^\nu h_{ij}^\nu)^2}{\sum_i h_{i1}^{\nu 2}}
f\left(\frac{M_{N_j}^2}{M_{N_1}}\right)\sin(\theta_1-\theta_j),
\end{equation}
where $f(x)=\sqrt{x}[1-(1+x)\ln\frac{1+x}{x}]$ and $\theta_j$ is given in eq.~(\ref{theta}). 
It is interesting that $CP$ phases which determine the $CP$ asymmetry $\varepsilon$
can be clearly traced in this model.
Since the neutrino oscillation data require $h_2$ and $h_3$ defined in eq.~(\ref{ny}) to 
be $O(10^{-3})$, $\varepsilon$ can be estimated as 
$\varepsilon=O(10^{-7})$ for $\rho_0\simeq\frac{\pi}{4}$.
This suggests that the lepton number asymmetry $\Delta L$ caused by this decay is 
given as $\Delta L=\varepsilon n_{N_1}^{\rm eq}=O(10^{-10})$ 
if the $N_1$ decay delays until the time when the washout of the generated lepton
number asymmetry is negligibly small.\footnote{We should remind the reader that such a situation 
can be realized for a sufficiently small $h_{i1}^\nu$ in a consistent way 
with the neutrino oscillation data.}
This $\Delta L$ is sufficient to give a required baryon number asymmetry through the 
sphaleron process. Since thermal leptogenesis could work successfully at a scale much smaller
than $10^8$ GeV, a lower bound of the reheating temperature estimated in the previous 
part is expected to be sufficient. 

\begin{figure}[t]
\begin{center}
\includegraphics[width=7.5cm]{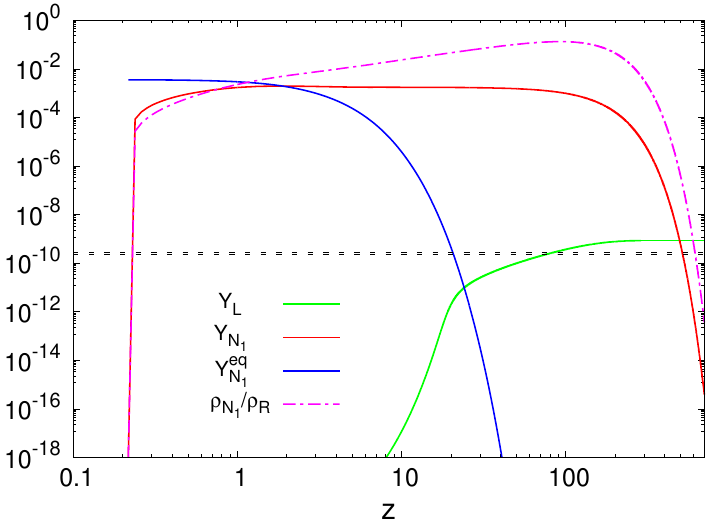}
\hspace{5mm}
\includegraphics[width=7.5cm]{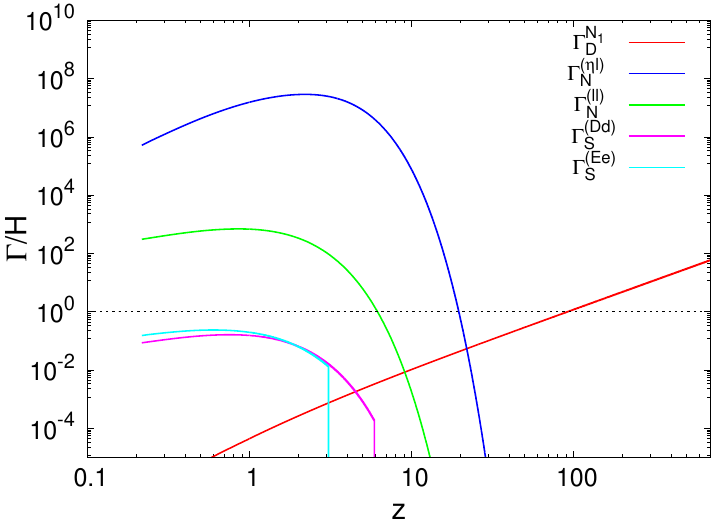}
\vspace*{-3mm}
\end{center}
\footnotesize{Fig. 2~~Left panel : Evolution of $Y_{N_1}$ and $Y_L\equiv|Y_\ell-Y_{\bar\ell}|$ 
as a function of $z(\equiv\frac{M_{N_1}}{T})$ starting from $z_R(\equiv\frac{M_{N_1}}{T_R})$. 
We set $Y_{N_1}(z_R)=Y_L(z_R)=0$ as initial conditions and the quantities given as the legend
are plotted. 
Horizontal dashed lines represent a region of $Y_L$ 
required to generate the observed baryon number asymmetry through the sphaleron 
process in the model.
Right panel : Evolution of the relevant reaction rate as a function of $z$. 
$\Gamma^{(ab)}_c$ stands for the reaction rate for the scattering $ab\rightarrow ij$ 
mediated by $c$ and $\Gamma_D^{N_1}$ is the decay width of $N_1$.}
\end{figure}

In order to examine it in a more quantitative way, we solve relevant Boltzmann equations 
numerically. 
We set parameters in the neutrino sector as
\begin{eqnarray}
&&y_N=(1.5\times 10^{-3}, 3\times 10^{-3}, 6\times 10^{-3}), \quad 
\tilde y_N=(1.5\times 10^{-3}, 0, 0), \nonumber \\
&& M_\eta=2~{\rm TeV}, \quad \tilde\lambda_5=10^{-5}, \quad h_1=2\times 10^{-8},
\end{eqnarray}
which gives $M_{N_1}=2121$~GeV, and then $M_{N_1}>M_{\eta^0}$ is satisfied.
For these parameters, the neutrino ocsillation data and eq.~(\ref{nmass}) fix 
the neutrino Yukawa coupling constants in eq.~(\ref{ny}) as
\begin{equation}
h_2=6.9\times 10^{-3}, \qquad h_3=2.3\times 10^{-3}.
\end{equation}
Using these and the parameters used in the previous examples,
we solve relevant Boltzmann equations for $Y_\psi(\equiv\frac{n_\psi}{s})$, where
$n_\psi$ is the number density of $\psi$ and $s$ is the entropy density \cite{ks}.
The result is shown in the left panel of Fig. 2, which proves that sufficient baryon 
number asymmetry $Y_B=3.0\times 10^{-10}$ is generated. 
In the right panel, the evolution of the reaction rates relevant to the Boltzmann equations
is plotted as a function of $z$. It shows that substantial decay of $N_1$ 
starts after the processes plotted as 
$\Gamma^{(\eta\ell)}_N$ and $\Gamma^{(\ell\ell)}_N$, which 
cause the washout of the lepton number asymmetry, are frozen out.
These figures support our above discussion on the leptogenesis in the present model.    
Even for the low reheating temperature estimated in the previous part,  
we find that thermal leptogenesis could occur successfully. 

\subsection{Electric dipole moment and $g-2$ of leptons}
New effects beyond the SM are expected to be caused radiatively  
by the additionally introduced fields. 
If we focus our study on ones in the lepton sector, 
the electric dipole moment of leptons is a typical example relevant to the $CP$ violation.
An operator relevant to it in the effective Lagrangian is given as
\begin{equation}
\frac{e c_{\alpha\beta}}{2\tilde m_\beta}\bar \psi_{L_\alpha} \sigma_{\mu\nu}
\psi_{R_\beta} F^{\mu\nu}+{\rm h.c.},
\label{effop}
\end{equation}
where $\psi_\alpha$ is a charged lepton mass eigenstate with mass $\tilde m_\alpha$.
It is related to the gauge eigenstate $\Psi_L=(\ell_L,E_L)^T$ through 
$\psi_L=\tilde V_L\Psi_L$ by using the unitary matrix $\tilde V_L$ defined in eq.(\ref{mass}).
The same operator also contributes to the anomalous magnetic moment of leptons and
lepton flavor violating processes such as $\ell_\beta \rightarrow \ell_\alpha\gamma$. 
Using the coefficient $c_{\alpha\beta}$ in eq.~(\ref{effop}),
new contributions to the electric dipole moment $d_{\psi_\alpha}$ of $\psi_\alpha$ and 
its anomalous magnetic moment 
$\Delta a_{\psi_\alpha}$  are represented as
\begin{equation}
d_{\psi_\alpha}=-\frac{e}{\tilde m_\alpha}{\rm Im}(c_{\alpha\alpha}), \qquad 
\Delta a_{\psi_\alpha}=2{\rm Re}(c_{\alpha\alpha}). 
\end{equation}
The branching ratio of the lepton flavor violating decay 
$\ell_\beta \rightarrow \ell_\alpha\gamma$ for the case $\tilde m_\beta \gg\tilde m_\alpha$ 
is also expressed  by using $c_{\alpha\beta}$ as
\begin{equation}
Br=\frac{48\pi^3\alpha_e}{({\tilde m_\alpha}^2G_F)^2}(|c_{\alpha\beta}|^2+
|c_{\beta\alpha}|^2),
\end{equation}
where $G_F$ is the Fermi constant and $\alpha_e$ is the fine structure constant of
the electromagnetic interaction.

One-loop diagrams contributing to this operator in the model are classified into three types
whose internal lines are composed of 
(i)~$E_{L,R}$ and a scalar $S$ or $\phi$,
(ii)~$N_{R_j}$ and $\eta$, and   
(iii)~$E_{L,R}$ and a $Z$ boson.
The formula for the coefficient $c_{\alpha\beta}$ caused by each diagram is presented 
in Appendix B. 
Here, we have to remind the reader that vector-like fermions are introduced to explain 
the $CP$ phases in the CKM and PMNS matrices in this model. 
This point is largely different from the models with vector-like leptons studied in \cite{veclep}. 
As a result, their effect on $c_{\alpha\beta}$ is expected to be largely suppressed 
since relevant off-diagonal components of the mixing matrix $\tilde V_L$ should 
be small enough to keep the approximate unitarity of the CKM and PMNS matrices \cite{pdg}. 

If we apply the parameters used in the previous parts to this calculation, we obtain the
predictions for the electric dipole moment as
\begin{eqnarray}
d_e=1.7\times 10^{-33},     \qquad d_\mu=4.6\times 10^{-29},    
\end{eqnarray}
where an $e\cdot{\rm cm}$ unit is used. A dominant contribution comes from the 
graph in type (i). 
These are much smaller than the present experimental upper bounds \cite{pdg}.  
The predicted anomalous magnetic moment of the electron and the muon is, respectively,
\begin{equation}
\Delta a_e=7.2\times 10^{-22}, \qquad \Delta a_\mu=1.2\times 10^{-15}.
\end{equation}
This shows that the muon anomalous magnetic moment reported at FNAL \cite{mug2} 
cannot be explained in this extended model.
On the lepton flavor violating decay $\mu\rightarrow e\gamma$, the branching ratio is 
predicted as
\begin{equation}
{\rm Br}(\mu\rightarrow e\gamma)= 1.4\times 10^{-21},
\end{equation}
which is also much smaller than the present bound \cite{mueg}.
These results show that it is difficult to find evidence of the model by using 
near future experiments for them.

\section{Summary} 
The SM has several issues for the $CP$ symmetry. 
Spontaneous $CP$ violation might give both a unified description for them and
a clue to study physics beyond the SM.  In this paper, on the basis of this point of view, 
we consider a model which could give a unified explanation for the $CP$ issues 
in the SM and study phenomenological consequences of the model.
The model is a simple extension of the SM with some fields including vector-like fermions 
and singlet scalars.  Since the model is constructed to be reduced to the scotogenic 
neutrino mass model at the low energy regions, it can also explain the small neutrino mass 
and the existence of DM in addition to the $CP$ issues. 

This model brings about the $CP$ phases in the CKM and PMNS matrices through 
the mixing between the ordinary fermions and the introduced vector-like fermions 
as a result of the spontaneous $CP$ violation in the scalar sector.
In the quark sector, since both contributions to $\bar\theta$ from radiative effects 
and higher order operators after the spontaneous $CP$ violation can be sufficiently 
suppressed, the strong $CP$ problem does not appear even if they are taken into account.
We also show that the model can cause a sufficient $CP$ asymmetry in the 
decay of the right-handed neutrinos and then the required baryon number asymmetry can be 
generated through low scale thermal leptogenesis. 

In order to show that the model works well, we present examples of parameter sets 
which realize rather good agreement with the CKM and PMNS matrices predicted through 
the various experimental results.    
Using these parameters, we prove that the observed baryon number asymmetry 
can be induced through thermal leptogenesis.
An interesting point in the leptogenesis is that the right-handed neutrinos can be 
produced sufficiently through the built-in interaction independently of the neutrino 
Yukawa couplings. 
As a result, the low scale leptogenesis occurs successfully in a consistent way with 
the neutrino oscillation data even if the mass of the right-handed 
neutrinos is of order of a TeV scale. 
It allows an inflation scenario in which the reheating temperature is of $O(10)$ TeV.
We present such an example of inflation which could be realized within the model. 

One-loop diagrams caused by the vector-like leptons in the model could contribute to 
the electric dipole moment and the anomalous magnetic moment of leptons and also
lepton flavor violating processes like $\mu\rightarrow e\gamma$.
However, since the vector-like leptons are introduced to explain the $CP$ phases 
in the PMNS matrix, its unitarity constraint heavily suppresses their effects to them.  
A similar feature is expected in the quark sector. Unfortunately,
it seems to be difficult to examine the model by observing them in 
near future experiments.

\newpage 
\section*{Apendix A ~Possible inflation in the model}
In this appendix, we discuss a possible inflation scenario in the model.
We suppose that the singlet scalar $S$ couples with the Ricci scalar in the Jordan frame as
\begin{eqnarray}
S_J = \int d^4x\sqrt{-g} \left[-\frac{1}{2}M_{\rm pl}^2R -
\xi_{S_1} S^\dagger S R -\frac{\xi_{S_2}}{2}(S^2+S^{\dagger 2})R
+\partial^\mu S^\dagger \partial_\mu S - V_0(S, S^\dagger) \right],
\end{eqnarray}
where $M_{\rm pl}$ is the reduced Planck mass. Its nonminimal couplings can be rewritten as
\begin{equation}
\frac{1}{2}\left[(\xi_{S_1}+\xi_{S_2})S_R^2+(\xi_{S_1}-\xi_{S_2})S_I^2\right]R,
\end{equation}
where $S_R$ and $S_I$ are real and imaginary parts of $S$, respectively, and 
defined as $S=\frac{1}{\sqrt 2}(S_R+iS_I)$.
We focus our consideration on a case where only one component $S_I$ is allowed to have 
the nonminimal coupling \cite{sinf2}. It can be realized by assuming a certain condition 
for $\xi_{S_1}$ and $\xi_{S_2}$ such as $\xi_{S_1}=-\xi_{S_2}$ and then it reduces to
an inflation model with  $\frac{1}{2}\xi S_I^2R$ where $\xi$ is fixed 
as $\xi\equiv \xi_{S_1}-\xi_{S_2}>0$. We review this scenario briefly here. 

If we consider the conformal transformation for a metric tensor in the Jordan frame
\begin{equation}
\tilde g_{\mu\nu}=\Omega^2g_{\mu\nu}, \qquad
\Omega^2=1+\xi\frac{S_I^2}{M_{\rm pl}^2},
\end{equation}
we have the action in the Einstein frame where the Ricci scalar term takes a canonical form
\cite{sinf1},
\begin{eqnarray}
S_E &=& \int d^4x\sqrt{-\tilde g} \Big[-\frac{1}{2}M_{\rm pl}^2\tilde R
+ \frac{1}{\Omega^2}\partial^\mu S_R \partial_\mu S_R +
\frac{1}{\Omega^4}\left(\Omega^2+6\xi^2 \frac{S_I^2}{M_{\rm pl}^2}\right) 
\partial^\mu S_I \partial_\mu S_I  \nonumber \\
&-&\frac{1}{\Omega^4}V_0(S_R, S_I) \Big],
\label{inflag}
\end{eqnarray}
where $V_0$ stands for the $\tilde\kappa_S$ term in eq.~(\ref{potv}).
We neglect $u$ in $V_0$ since it is much smaller than $O(M_{\rm pl})$ 
that is a value of $S_I$ during the inflation. 
The kinetic term of $S_I$ in eq.~(\ref{inflag}) can be rewritten to the canonical form
by inflaton $\chi_c$ which is defined by 
\begin{equation}
\Omega^2\frac{d\chi_c}{dS_I}=\sqrt{\Omega^2+6\xi^2\frac{S_I^2}{M_{\rm pl}^2}}.  
\label{tchi}
\end{equation}
The potential of $\chi_c$ can be fixed through $V(\chi_c)=\frac{1}{\Omega^4}V(S_I)$ 
by using this relation. It can be approximately expressed as 
$V=\frac{\tilde\kappa_S}{4\xi^2}M_{\rm pl}^4$ at the large field regions $\chi_c>M_{\rm pl}$.
Results of the CMB observations put constraints on the model parameters in 
the potential $V$.
The slow-roll parameters in this model can be evaluated by using 
eq.~(\ref{tchi}) as \cite{inf,inf1}
\begin{equation}
\epsilon\equiv\frac{M_{\rm pl}^2}{2}\left(\frac{V^\prime}{V}\right)^2
=\frac{8M_{\rm pl}^4}{\xi\left(1+6\xi\right)\chi_c^4}, \qquad
\eta\equiv M_{\rm pl}^2\frac{V^{\prime\prime}}{V}=
-\frac{8M_{\rm pl}^2}{\left(1+6\xi\right)\chi_c^2},
\end{equation}
where $V^\prime$ stands for $\frac{dV}{d\chi_c}$.
If we use the $e$-foldings number ${\cal N}_k$ from the time when the scale 
$k$ exits the horizon to the end of inflation, these slow-roll parameters are 
found to be approximated as $\epsilon\simeq \frac{3}{4{\cal N}_k^2}$ and 
$\eta\simeq -\frac{1}{{\cal N}_k}$.
Thus, the model predicts favorable values for the scalar power index as 
$n_s=0.958-0.965$ and the tensor to
scalar ratio as $r=0.0048-0.0033$ for ${\cal N}_k=50-60$.

The spectrum of the CMB density perturbation predicted by the slow roll inflation 
is known to be expressed as \cite{inf,inf1}
\begin{equation}
{\cal P}(k)=A_s\left(\frac{k}{k_\ast}\right)^{n_s-1},  \qquad
A_s=\frac{V}{24\pi^2M_{\rm pl}^4\epsilon}\Big|_{k_\ast}. 
\label{power}
\end{equation}
If we use the Planck data $A_s=(2.101^{+0.031}_{-0.034})\times 10^{-9}$ 
at $k_\ast=0.05~{\rm Mpc}^{-1}$ \cite{planck18}, we find a constraint on the coupling
constant $\tilde\kappa_S$ as 
\begin{equation}
\tilde\kappa_S\simeq 1.2\times 10^{-6}\left(\frac{\xi}{50}\right)^2 
\left(\frac{55}{{\cal N}_{k_\ast}}\right)^2,
\label{kap}
\end{equation}
and the Hubble parameter satisfies $H_I=1.5\times 10^{13}
\left(\frac{55}{{\cal N}_{k_\ast}}\right)$~GeV during the inflation. 

\section*{Apendix B ~Formulas for radiative processes in the lepton sector}
In this appendix, we present formulas of the coefficient $c_{\alpha\beta}$ in 
eq.~(\ref{effop}) caused by one-loop diagrams \cite{tmueg}.
Diagrams of types (i) and (ii) are shown in Fig.~3. 
Yukawa interactions relevant to (i) are given in eq.~(\ref{llag}).   
They are expressed by using the mass eigenstates $\psi_\alpha$ as 
\begin{eqnarray}
&&\sum_{\alpha,\beta,a=1}^4\left[
\left(\sum_{j=1}^3\frac{x_j}{\sqrt 2}
\tilde V^{L}_{\alpha j}\right) \tilde V^{R\dagger }_{4\beta}O^T_{1a}
\chi_a\bar\psi_{L\alpha} \psi_{R\beta} + 
\left(\sum_{j=1}^3\frac{y_j^e+\tilde y_j^e}{\sqrt 2}
 \tilde V^{R\dagger}_{j\beta }\right) \tilde V^{L}_{4\alpha}O^T_{2a}
\chi_a\bar\psi_{L\alpha} \psi_{R\beta}\right. \nonumber \\
&&+\left.\left(\sum_{j=1}^3\frac{i(y_j^e-\tilde y_j^e)}{\sqrt 2}
\tilde V^{R\dagger}_{j\beta}\right)\tilde V^{L}_{4\alpha}O^T_{3a}
\chi_a\bar\psi_{L\alpha} \psi_{R\beta} +{\rm h.c.}\right],
\end{eqnarray}
where $\tilde V^R$ is a unitary matrix which diagonalizes 
the lepton mass matrix ${\cal M}_\ell$ as 
$\tilde V^{L}{\cal M}_\ell\tilde V^{R\dagger}={\cal M}_\ell^{\rm diag}$.
Taking account of these interactions, the contribution from these diagrams 
to $c_{\alpha\beta}$ can 
be calculated as
\begin{eqnarray}
c_{\alpha\beta}^{S\phi}&=&\frac{1}{16\pi^2}
\left(\sum_{i=1}^3x_i\tilde V_{i\alpha}^{L}\right)
\left(\sum_{j=1}^3(y_j+\tilde y_j)\tilde V_{j\beta}^{R\dagger}\right)
\sum_{a,\gamma=1}^4O^T_{1a}O^T_{2a}\tilde  V_{4\gamma}^L\tilde V_{\gamma 4}^{R \dagger}
\frac{\tilde m_{\gamma}\tilde m_\beta}{m_a^2}
J\left(\frac{\tilde m_{\gamma}^2}{m_a^2}\right) \nonumber\\
&+&\frac{i}{16\pi^2}
\left(\sum_{i=1}^3x_i\tilde V_{i\alpha}^{L}\right)
\left(\sum_{j=1}^3(y_j-\tilde y_j)\tilde V_{j\beta}^{R\dagger}\right)
\sum_{a,\gamma=1}^4O^T_{1a}O^T_{3a}\tilde V_{4\gamma}^{L}\tilde  V_{\gamma 4}^{R\dagger}
\frac{\tilde m_{\gamma}\tilde m_\beta}{m_a^2}
J\left(\frac{\tilde m_{\gamma}^2}{m_a^2}\right), \nonumber \\
c_{\alpha\beta}^S&=&\frac{1}{32\pi^2} 
\left(\sum_{i=1}^3(y_i+\tilde y_i)\tilde V_{i \alpha }^{R}\right)
\left(\sum_{j=1}^3(y_j+\tilde y_j)\tilde V_{j\beta}^{R\dagger}\right)
\sum_{a,\gamma=1}^4O^T_{2a}O^T_{2a}
\tilde V_{4\gamma}^{L \dagger}\tilde  V_{\gamma 4}^L
\frac{(\tilde m_\alpha+\tilde m_\beta)\tilde m_{\beta}}{2m_a^2}
H\left(\frac{\tilde m_{\gamma}^2}{m_a^2}\right)\nonumber \\
&+&\frac{1}{32\pi^2}
\left(\sum_{i=1}^3(y_i-\tilde y_i)\tilde V_{i \alpha }^{R}\right)
\left(\sum_{j=1}^3(y_j-\tilde y_j)\tilde V_{j\beta}^{R\dagger}\right)
\sum_{a,\gamma=1}^4O^T_{3a}O^T_{3a}
\tilde V_{4\gamma}^{L \dagger} \tilde V_{\gamma 4}^L
\frac{(\tilde m_\alpha+\tilde m_\beta)\tilde m_\beta}{2m_a^2}
H\left(\frac{\tilde m_{\gamma}^2}{m_a^2}\right), \nonumber\\ 
c_{\alpha\beta}^{\phi}&=&\frac{1}{32\pi^2}
\left(\sum_{i=1}^3x_i\tilde V_{i\alpha}^{L}\right)\left(\sum_{j=1}^3x_j \tilde V_{j\beta}^{L\dagger}\right)
\sum_{a,\gamma=1}^4 O^T_{1a}O^T_{1a}\tilde V_{4\gamma}^{R\dagger}\tilde V_{\gamma 4}^R
\frac{(\tilde m_\alpha+\tilde m_{\beta})\tilde m_\beta}{2m_a^2}
H\left(\frac{\tilde m_\gamma^2}{m_a^2}\right), 
\end{eqnarray}
where $\tilde m_{4}=\tilde M_E$ and 
$m_a^2$ is the $a$th eigenvalue of the mass matrix ${\cal M}_s^2$ and 
given in eq.~(\ref{smass}).  $c_{\alpha\beta}^{S\phi}$,
$c_{\alpha\beta}^S$, and $c_{\alpha\beta}^\phi$ are contributions caused by the left three diagrams 
shown in Fig.~3, respectively. Loop functions $J(r)$ and $H(r)$ are defined as 
\begin{eqnarray}
&&J(r)=\frac{1}{2(r-1)^3}(3-4r+r^2+ 2\ln r), \nonumber \\  
&& H(r)=\frac{1}{6(r-1)^4}(2+3r-6r^2+r^3+6r\ln r).
\end{eqnarray}
One might expect that the diagram in which the chirality flip occurs in the internal fermion line 
brings about the enhancement via its large mass. However,
since $\tilde V_L$ is related to the PMNS matrix in this model,
the unitarity requirement makes the mixing between the light leptons and 
vector-like leptons small. As a result, effective coupling is considered to be strongly 
suppressed, and the enhancement is ineffective.   

\begin{figure}[t]
\begin{center}
\includegraphics[width=15cm]{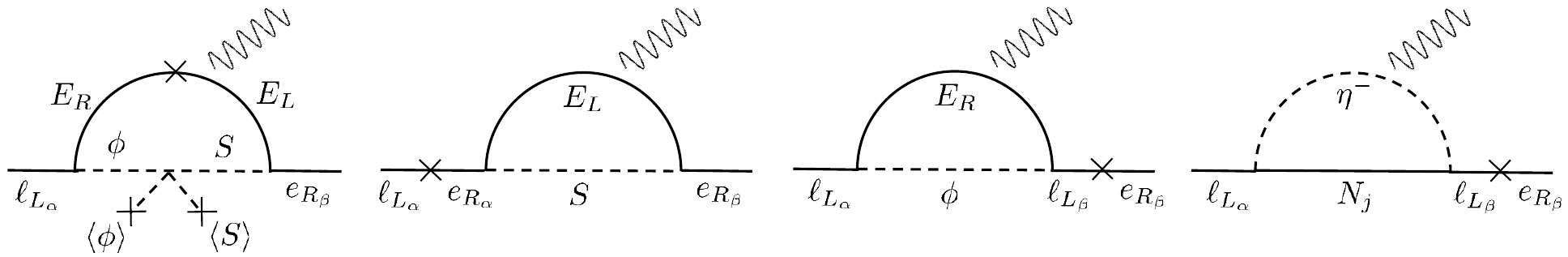}
\vspace*{-2mm}
\end{center}
\footnotesize{Fig.~3~~One-loop diagrams caused by the scalar 
exchange, which give new contribution to the effective 
operator in eq.~(\ref{effop}). They are drawn by using the gauge eigenstates. }
\end{figure}

Yukawa interactions relevant to (ii) are given in eq.~(\ref{nlag}). 
If we rewrite them by using the mass eigenstates $\psi_\alpha$,
they can be expressed as
\begin{equation}
h_{ij}^\nu\tilde V_{\alpha i}^{L}\bar\psi_{L_\alpha}\eta^- N_j + {\rm h.c.}. 
\end{equation}
Their contribution to the coefficient $c_{\alpha\beta}$ can be calculated as
\begin{equation}
c_{\alpha\beta}^\eta=\frac{1}{ 16\pi^2}
\sum_{i,j,k=1}^3h^\nu_{ik}\tilde V_{\alpha i}^Lh^\nu_{jk}\tilde V_{\beta j}^L e^{i\theta_k }
\frac{(\tilde m_\alpha+\tilde m_\beta)\tilde m_\beta}{2m_0^2}I\left(\frac{M_{N_k}^2}{m_0^2}\right), 
\end{equation}
where $M_{N_k}$ and $\theta_k$ are given in eq.~(\ref{theta}). A loop function $I(r)$ is defined 
as 
\begin{equation}
I(r)=\frac{1}{6(r-1)^4}(-1+6r-3r^2-2r^3+6r^2\ln r).
\end{equation}
If we apply tribimaximal assumption (\ref{ny}) to this formula,
coefficients relevant to interesting quantities can be rewritten as
\begin{eqnarray}
c_{11}^\eta&=&\frac{1}{16\pi^2}\frac{\tilde m_1^2}{m_0^2}h_3^2e^{i\theta_3}
I\left(\frac{M_{N_3}^2}{m_0^2}\right),
\nonumber \\
c_{22}^\eta&=&\frac{1}{16\pi^2}\frac{\tilde m_2^2}{m_0^2}\left[h_1^2e^{i\theta_1}I
\left(\frac{M_{N_1}^2}{m_0^2}\right)
+h_2^2e^{i\theta_2}I\left(\frac{M_{N_2}^2}{m_0^2}\right)
+h_3^2e^{i\theta_3}I\left(\frac{M_{N_3}^2}{m_0^2}\right)\right], \nonumber \\
c_{12}^\eta&=&\frac{1}{32\pi^2}\frac{\tilde m_2^2}{m_0^2}h_3^2e^{i\theta_3}
I\left(\frac{M_{N_3}^2}{m_0^2}\right),
\end{eqnarray}
where we use the assumption that $\tilde V^L$ is almost diagonal.

\begin{figure}[t]
\begin{center}
\includegraphics[width=10cm]{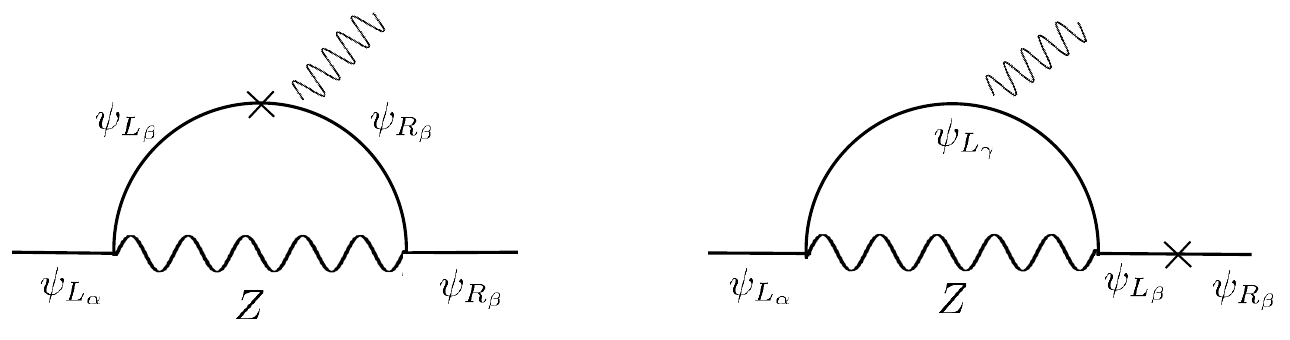}
\vspace*{-2mm}
\end{center}
\footnotesize{Fig.~4~~One-loop diagrams caused by the $Z$ boson exchange, 
which give the effective operator in eq.~(\ref{effop}) deviated from the SM one.
They are drawn by using the mass eigenstates.}
\end{figure}

Diagrams of type (iii) are shown in Fig.~4. 
Gauge interaction relevant to these is given as
\begin{equation}
\frac{g}{\cos\theta_W}\left[\sum_{j=1}^3\left(g_L\bar\ell_{L_j}\gamma^\mu\ell_{L_j}+
g_R\bar e_{R_j}\gamma^\mu e_{R_j}\right)+
g_R\bar E_L\gamma^\mu E_L+ g_R\bar E_R\gamma^\mu E_R \right] Z_\mu,
\end{equation}
where $g_L$ and $g_R$ are defined as 
$g_L=-\frac{1}{2}+\sin^2\theta_W$ and $g_R=\sin^2\theta_W$.
Since the extra lepton $E_L$ is introduced as an $SU(2)_L$ singlet, flavor changing couplings appear 
only in the left-handed neutral current part as
\begin{equation}
\frac{g}{\cos\theta_W}\sum_{\alpha=1}^4\left[\sum_{\beta=1}^4\bar\psi_{L_\alpha}\gamma^\mu
({\cal C}_L)_{\alpha\beta}\psi_{L_\beta} +g_R\bar\psi_{R_\alpha}
\gamma^\mu\psi_{R_\alpha}\right] Z_\mu,
\label{nc}
\end{equation}
where a charge matrix ${\cal C}_L$ is expressed as ${\cal C}_L=\tilde V_L C_L\tilde V_L^\dagger$.
Although $C_L$ is a diagonal matrix,  its elements are $(g_L, g_L, g_L, g_R)$, and then
${\cal C}_L$ has nonzero off-diagonal components to cause flavor mixings.
Their contribution to $c_{\alpha\beta}$ can be calculated as
\begin{eqnarray}
c_{\alpha\beta}^Z&=&\frac{1}{16\pi^2}\frac{g^2}{\cos^2\theta_W}
({\cal C}_L)_{\alpha\beta}g_R
\frac{\tilde m_{\beta}^2}{m_Z^2}F\left(\frac{\tilde m_{\beta}^2}{m_Z^2}\right)  \nonumber \\
&+&\frac{1}{32\pi^2}\frac{g^2}{\cos^2\theta_W}
\sum_{\gamma=1}^4({\cal C}_L)_{\alpha\gamma}
({\cal C}_L)_{\gamma\beta}\frac{(\tilde m_\alpha+\tilde m_\beta)\tilde m_{\beta}}{2m_Z^2}
G\left(\frac{\tilde m_{\gamma}^2}{m_Z^2}\right) - c_{SM}\delta_{\alpha\beta} ,
\label{zcont}
\end{eqnarray}
where $c_{SM}$ represents the contribution in the SM corresponding to other two terms. 
$F(r)$ and $G(r)$ are loop functions defined as
\begin{eqnarray}
&&F(r)=\frac{1}{2(r-1)^3}(-4+3r +r^3-6r\ln r), \nonumber \\
&&G(r)=\frac{1}{6(r-1)^4}(-8+38r-39r^2+14r^3-5r^4+18r^2\ln r).
\end{eqnarray}
Although the chirality flip occurs in the internal line in the first diagram,  
the enhancement via large fermion mass is not caused since the right-handed
current is flavor diagonal.  

\newpage
\bibliographystyle{unsrt}

\end{document}